# Effect of phonon anharmonicity on ferroelectricity in $Eu_xBa_{1-x}TiO_3$

Bommareddy Poojitha,[1,†] Km Rubi,[2,†] Soumya Sarkar,[3,4,†] R. Mahendiran,[2,3,*]
T. Venkatesan,[2,3,4,*] and Surajit Saha,[1,*]

[1]Department of Physics, Indian Institute of Science Education and Research, Bhopal 462066, INDIA

[2]Department of Physics, National University of Singapore, Singapore 117551, SINGAPORE

[3]NUSNNI-NanoCore, National University of Singapore, Singapore 117411, SINGAPORE

[4]NUS Graduate School for Integrative Sciences and Engineering, Singapore 117456, SINGAPORE

*Authors to whom correspondence to be addressed: phyrm@nus.edu.sg, venky@nus.edu.sg, surajit@iiserb.ac.in

**Investigating the competition between ferroelectric ordering and quantum fluctuations is essential to tailor the desired functionalities of mixed ferroelectric and incipient ferroelectric systems, like, (Ba,Sr)TiO$_3$ and (Eu,Ba)TiO$_3$. Recently, it has been shown that suppression of quantum fluctuations increases ferroelectric ordering in (Eu,Ba)TiO$_3$ and since these phenomena are coupled to crystallographic phase transitions it is essential to understand the role of phonons. Here, we observe that the unusual temperature dependence of phonons in BaTiO$_3$ gets suppressed when Ba$^{2+}$ is replaced by Eu$^{2+}$. This manifests in a decrease in the cubic-to-tetragonal (i.e., para-to-ferroelectric) phase transition temperature (by 150 K) and a complete suppression of tetragonality of the lattice (at room temperature by 40% replacement of Ba$^{2+}$ by Eu$^{2+}$). We have quantified the anharmonicity of the phonons and observed that the replacement of Ba$^{2+}$ by Eu$^{2+}$ suppresses it (by 93%) with a resultant lowering of the ferroelectric ordering temperature in the Eu$_x$Ba$_{1-x}$TiO$_3$. This suggests that tuning phonon anharmonicity can be an important route to novel ferroelectric materials.**

## INTRODUCTION:

Ferroelectric materials are characterized by features like spontaneous polarization, nonlinear dielectric behavior, large electro-optic coefficient, high permittivity, as well as pyroelectric and piezoelectric properties, making it possible to design various functional devices out of these materials [1]. One of the most studied ferroelectric materials with perovskite structure is $BaTiO_3$, which also exhibits a rich variety of structural phases. $BaTiO_3$ undergoes a series of structural transitions with increasing temperature [2], showing a low temperature rhombohedral to orthorhombic phase transition at around 193 K. This is followed by an orthorhombic to tetragonal phase change at around 280 K and then a tetragonal to cubic transition at around 395 K. $BaTiO_3$ is paraelectric in the high temperature cubic phase. However, in the low temperature phases it becomes a ferroelectric as the macroscopic electric polarization aligns in parallel to [001], [011], and [111] axes (denoted in terms of cubic notation) of the thermally-induced crystallographic phases characterized by tetragonal, orthorhombic, and rhombohedral unit cell, respectively [2]. On the contrary, $EuTiO_3$ (also a perovskite) is an incipient ferroelectric where there exists a competition between the ferroelectric order parameter and quantum fluctuations [3,4]. Though $EuTiO_3$ has a soft polar phonon mode, the polarization does not freeze down to the lowest temperature due to quantum fluctuations [4]. Further, unlike the other similar incipient ferroelectrics e.g. $SrTiO_3$ and $KTaO_3$, $EuTiO_3$ undergoes a paramagnetic to G-type antiferromagnetic transition below about 5 K due to its localized 4f electrons (S=7/2) [5] and exhibits a giant magnetodielectric effect [3,4].

Inducing ferroelectric ordering in incipient ferroelectrics has been of great importance not only being fundamentally interesting but also because it is important for realizing their technological potentials. Introduction of $Ba^{2+}$ in $SrTiO_3$ and $Nb^{5+}$ in $KTaO_3$ have been known to induce ferroelectric ordering [6-8]. Recently, it was demonstrated that by doping $Ba^{2+}$ in $EuTiO_3$ one could suppress the quantum fluctuations in $EuTiO_3$, thereby inducing ferroelectric ordering [9]. Besides the suppression of quantum fluctuations, magnetoelectric effects have also been observed in $Ba^{2+}$ doped $EuTiO_3$ [3,10,11]. Moreover, (Eu,Ba)$TiO_3$ has been proposed as a model system to search for electron electric dipole moment which is of vital importance to study Charge-Parity (CP) violation [12].

Phonons have an intimate relationship with ferroelectric ordering in $BaTiO_3$ and so should be the case in the incipient ferroelectric $EuTiO_3$ upon doping $Ba^{2+}$. However, the exact mechanism by which phonons play a role in this competition between ferroelectric ordering and quantum fluctuations remains unclear. Notably, in such incipient ferroelectrics (like $EuTiO_3$, $SrTiO_3$ etc.) the phase transitions are coupled to lattice, causing these materials to exhibit quantum criticality over a wide range of temperatures [13]. At such finite temperatures phonon anharmonicity, (*i.e.*, phonon-phonon scattering) becomes crucial for understanding the ground state of the incipient ferroelectrics [3,14]. In this article, we investigate the phonons anharmonicities in $BaTiO_3$ as a function of $Eu^{2+}$ doping in order to establish the correlation between anharmonicity and ferroelectric ordering in the $Eu_xBa_{1-x}TiO_3$ systems.

**RESULTS AND DISCUSSION:**

*Structure and phase purity of $Eu_xBa_{1-x}TiO_3$* - Polycrystalline $BaTiO_3$ and $Eu_xBa_{1-x}TiO_3$ (x = 0.1, 0.2, 0.3, 0.4, and 1) powder samples were synthesized using a solid state reaction route (see Sec. III) and are found to be in a highly pure phase (see Sec. S1 and Figs. S1 to S3 in the supplemental material for details). The rich variety of structural phases exhibited by these samples have been captured in the temperature-dependent x-ray diffraction patterns, shown in Figs 1(a)-1(e). Figure 1(c) shows the changes in the lattice parameters as a function of temperature in the four different structural phases [15]. A partial replacement of $Ba^{2+}$ by $Eu^{2+}$ (such as in $Eu_{0.2}Ba_{0.8}TiO_3$) reduces the lattice parameters as shown in Figure 1(f) (and Figure S2 in supplemental material) and lowers the cubic-to-tetragonal phase transition temperature. However, no significant change in the transition temperatures of the low temperature phases could be detected with $Eu^{2+}$ doping whose reason is not clear at present. However, as the para-to-ferroelectric phase transition is specifically associated with the cubic-to-tetragonal structural change, our emphasis in this article will be on this crystallographic phase transition.

*Γ-point Raman phonons of $Eu_xBa_{1-x}TiO_3$* -The ferroelectric transition in $BaTiO_3$ is ferrodistortive in nature. Therefore, in both the ferro- and para-electric phases it has one formula unit (5 atoms) per unit cell that gives rise to twelve optical phonon modes at the Γ-point of the Brillouin zone. In the paraelectric-cubic phase none of the phonons ($3F_{1u}+F_{2u}$) are Raman active whereas in the ferroelectric-tetragonal phase there are $3A_1$, $B_1$, and $4E$ Raman modes. A detailed description of the origin of these Raman modes at the Γ-point are provided in the supplemental material (Section S2). Figure 2 (a,b) shows the typical Raman spectra presenting the prominent phonon modes of $BaTiO_3$ and $Eu_{0.2}Ba_{0.8}TiO_3$ at 78 K (obtained via Lorentz spectral fitting). The observed mode frequencies and their assignments based on previous reports [16-21] are listed in Table-1. As can be seen in Table-1, the prominent modes near 265 cm$^{-1}$ (*P5*) and 520 cm$^{-1}$ (*P8*) undergo a *red-shift* upon $Eu^{2+}$ incorporation which can be attributed to an increase in the atomic mass while the *P6* mode at 308 cm$^{-1}$ does not respond to the change in atomic mass because Ba/Eu atoms are not involved in this vibration [21]. Figure S4 (Supplemental material) shows the dependence of the Raman spectrum at 300 K as a function of $Eu^{2+}$ incorporation by replacing $Ba^{2+}$ in $BaTiO_3$. Absence of the 308 cm$^{-1}$ (*P6*) mode for 40% doping of $Eu^{2+}$ in $BaTiO_3$ (i.e. $Eu_{0.4}Ba_{0.6}TiO_3$) and above suggests that its crystal structure is not tetragonal at 300 K [16,18,20-23] and, therefore, it is no longer a room temperature ferroelectric. Notably, the other modes show a small or almost no shift in phonon frequency at room temperature upon Eu-incorporation implying that there is an indirect or no involvement of Ba/Eu atoms in these vibrations. However, as shown by Freire and Katiyar [21] in their lattice dynamical calculations (LDC), all the vibrations at the Γ-point involve all the three atoms (Ba, Ti, O) excepting the $B_1$ mode at 308 cm$^{-1}$. Our data at room temperature, therefore, corroborate with the LDC for the prominent modes at 265 ($A_1$), 308 ($B_1$), and 520 ($A_1$) cm$^{-1}$. However, the corresponding '*red-shift*' for the other modes due to Eu-incorporation is not observed, which is highly unusual. In order to understand this phenomenon, we have performed a temperature-dependent study which is discussed below. The shaded region of the spectra below 100 cm$^{-1}$ is influenced by the cut-off of the band-pass filter used during the measurements and, hence, will not be used for drawing any conclusion.

Figure 2(c,d) shows the temperature-dependent Raman spectra of BaTiO$_3$ and Eu$_{0.2}$Ba$_{0.8}$TiO$_3$ to emphasize the effect of structural changes on the Raman spectrum (temperature dependence of the Raman spectra of Eu$_x$Ba$_{1-x}$TiO$_3$ for x = 0.1, 0.3, 0.4, and 1 are shown in Figures S5 and S6 in supplemental material). For BaTiO$_3$, it may be noted that as the temperature increases there is a clear and systematic rise in the background spectral line shape below 200 cm$^{-1}$. Such a line shape at lower frequencies has been reported previously [16,19-21,24-26] and may be attributed to the contribution from the soft $E$(TO$_1$) mode at 35 cm$^{-1}$ (not captured in our data due to the cut-off of band-pass filter) [16,27]. A similar temperature-dependent change in the line shape at lower frequencies can also be seen in Eu$_{0.2}$Ba$_{0.8}$TiO$_3$ but occurs at even lower temperatures (Figure 2d) which may be associated with the phonon anharmonicity and a lower unit cell volume of Eu$_{0.2}$Ba$_{0.8}$TiO$_3$. Further, mode P6 at 308 cm$^{-1}$ is a representative of the ferroelectric phase of BaTiO$_3$ [16,20-23,25] that can be seen at temperatures below ~ 400 K (Figure 2c) whereas this mode in Eu$_{0.2}$Ba$_{0.8}$TiO$_3$ is absent above

~ 360 K (Figure 2d) thus suggesting a lowering of the para-to-ferroelectric transition temperature in $Eu_{0.2}Ba_{0.8}TiO_3$. Most noticeable changes in the spectra (of both $BaTiO_3$ and $Eu_{0.2}Ba_{0.8}TiO_3$) with varying temperature occur as (dis)appearance of phonon modes and unusual / anomalous shift of the modes. In order to quantify these temperature-dependent changes, the spectra at various temperatures have been fitted with Lorentzian functions identifying the various phonons (peaks P1 to P10 shown in Figure 2a and 2b) [17,21]. It must be noted that during the spectral fitting we have not considered the coupling of modes with $A_1$ symmetry as was proposed earlier [24,25] because: (a) our samples are polycrystalline powders where disorders are endemic and hence it is difficult to materialize such polariton couplings, and (b) we have not observed a very clear presence of the well known '*depolarization dip*' near 180 $cm^{-1}$ in the tetragonal phase thus signifying the absence of phonon coupling [25-29]. The temperature dependence of the phonon mode frequency ($\omega$) for both $BaTiO_3$ and $Eu_{0.2}Ba_{0.8}TiO_3$ shows characteristic signatures of the structural phase transitions (see Figure 3). Notably, the peaks P4 and P7 near 240 and 489 $cm^{-1}$, respectively, in $BaTiO_3$ disappear and a new peak (P9) appears at 552 $cm^{-1}$ above 190 K owing to the rhombohedral to orthorhombic phase transition. A '*jump*' [17] in the phonon frequency for the peaks P3 and P5 as well as a change in slope ($\omega$ vs T) for the peaks P1 and P8 can be observed across the structural transition. As the temperature increases beyond 280 K, the peaks P1 and P2 become weak, finally merging with the broad background (that arises from the low frequency $E(TO_1)$ mode). The peaks P3 and P5 show another '*jump*' in frequency while a change in slope ($\omega$ vs T) for P6 and P9 can be seen at the same temperature which can be attributed to the orthorhombic to tetragonal phase transition at ~ 280 K. It should be noted that though for most of the peaks the '$\omega$' shows a decreasing trend with increasing temperature which is a signature of normal lattice expansion due to quasi-harmonic behaviour, the peaks P2, P9, and P10 show an anomalous trend (an increase in frequency) with temperature. Such an anomaly may arise from strong phonon-phonon anharmonic interactions [30-33] which will be discussed later. The behaviour of the respective phonon modes in Eu$_{0.2}$Ba$_{0.8}$TiO$_3$, with regard to their appearance/disappearance at crystallographic phase transition temperatures is similar to that in BaTiO$_3$ (see Figure 2d).

Notably, Eu doping also gives rise to two stark differences in the temperature-dependent evolution of the phonon modes in BaTiO$_3$. First, the temperature-dependent shift of most phonon modes in BaTiO$_3$ are suppressed upon Eu doping (the largest being for the P5 mode ~ 25 cm$^{-1}$). Moreover, the phonon modes in the tetragonal crystal structure disappear at a 40 K lower temperature in Eu$_{0.2}$BaTiO$_3$ as compared to BaTiO$_3$. In addition, the '*jump*' in frequency across the structural transitions is not observed in P3 and P5.

The difference in the temperature dependence (in all the phases) of the modes (Figure 3) upon incorporating Eu$^{2+}$ (by replacing Ba$^{2+}$) in BaTiO$_3$ could be due to the effect of: (*i*) change in mass and (*ii*) change in lattice (local symmetries) both of which would lead to the changes in phonon anharmonicities. Replacement of lighter Ba$^{2+}$ ions by heavier Eu$^{2+}$ ions in BaTiO$_3$ would *red-shift* the mode frequency (because $\omega \propto m^{-1/2}$) if Ba atom is involved in a vibration (all the modes involve Ba atoms except the *P6* mode at 308 cm$^{-1}$). On the other hand, the ionic radius of Eu$^{2+}$ is lower than that of Ba$^{2+}$ that shrinks the unit cell (at 300 K - BaTiO$_3$: a=b=3.9853 Å, c=4.0069 Å; Eu$_{0.2}$Ba$_{0.8}$TiO$_3$: a=b=3.9789 Å, c=3.9982 Å as shown in Figures 1 and S2), which is expected to *blue-shift* the mode frequencies. In other words, these two effects (change in mass and lattice) will compete and show a resultant effect on the mode frequencies and phonon anharmonicities. As can be seen in Figure 3, all the modes show a clear *red-shift* of frequency in Eu$_{0.2}$Ba$_{0.8}$TiO$_3$ as compared to that in BaTiO$_3$ except for P4, P7, P9, and P10 wherein a noticeable *blue-shift* can be observed. The temperature dependence of frequency of a phonon (*i*) may be expressed as [32-34]:

$$\omega^i(T) = \omega^i(0) + \Delta\omega^i_{qh}(T) + \Delta\omega^i_{anh}(T) + \Delta\omega^i_{el-ph} + \Delta\omega^i_{sp-ph} \quad (1)$$

where, $\omega^i(0)$ is the phonon frequency at 0 K. The term $\Delta\omega^i_{qh}(T)$ is the change in frequency due to the quasiharmonic contribution arising from the change in lattice volume (i.e., force constant) without changing the phonon population. The term $\Delta\omega^i_{anh}(T)$ is the change in frequency due to the intrinsic anharmonic contribution arising from the real part of the self-energy of a phonon decaying into two (cubic anharmonicity) or three (quartic anharmonicity) phonons. On the other hand, the terms $\Delta\omega^i_{el-ph}$ and $\Delta\omega^i_{sp-ph}$ are the changes in phonon frequency arising due to a coupling of the phonon with the charge carriers and spins, respectively, both of which are absent in these insulating and non-magnetic materials ($BaTiO_3$ and $Eu_{0.2}Ba_{0.8}TiO_3$). Therefore, we can attribute the observed changes in phonon frequencies to the phonon anharmonicities.

Temperature-dependent changes in phonon anharmonicity are not only related to lattice volume but also to the phonon population (phonon density). In fact, phonon anharmonicities become significant when the temperature increases causing a phonon to decay into two or three phonons owing to cubic or quartic anharmonic interactions (or higher order interactions at much higher temperatures). Considering the three-phonon process (cubic anharmonicity), the temperature dependence of a phonon frequency ($i$) can be expressed as [34]:

$$\omega^i(T) = \omega^i(0) + A\left[1 + \frac{2}{e^x - 1}\right] \qquad (2)$$

where, $x = \hbar\omega^i(0)/2k_B T$, $\hbar$ is reduced Planck's constant, $k_B$ is Boltzmann constant, $T$ is the temperature, and $A$ is the coefficient of anharmonicity. A comparison of the equations (1) and (2) indicates that the coefficient of anharmonicity $'A'$ weighs the contributions of quasiharmonic effect thus implicitly associating with the mode Grüneisen parameter and intrinsic anharmonic effects (see section S4 in supplemental material). As mentioned earlier, the most surprising behaviour is observed for the mode P5 in BaTiO$_3$ ($A_1$ mode at 265 cm$^{-1}$) that undergoes an unusually large change in frequency ~ 20 cm$^{-1}$ over the temperature range of 300 to 400 K (see Figure 3), indicating a strong phonon anharmonicity. Interestingly, upon replacing Ba$^{2+}$ by Eu$^{2+}$ the mode becomes nearly temperature-independent, as shown in Figure 3. To understand this behaviour of P5, we have quantified its phonon anharmonicity by fitting the frequency vs temperature using equation (2) (see Figure S7 in Supplemental material). As shown in Figure 4a, the mode P5 has a very high coefficient of anharmonicity in pure BaTiO$_3$ which decreases with increasing doping of Eu$^{2+}$, clearly indicating a suppression of phonon anharmonicity (by ~ 93 %) with increasing Eu$^{2+}$. The P6 mode ($B_1$ at 308 cm$^{-1}$), which is present only in the ferroelectric phase also exhibits a similar decrease in anharmonicity (by ~ 73%) with increasing substitution of Ba$^{2+}$ by Eu$^{2+}$ (as shown in Figure 4a). It is important to note that the atomic displacements have an important role in phonon anharmonicity and the participating atoms in modes P5 and P6 displace along the crystallographic c-axis, shown in Figure 4(b), which decides the tetragonality of the lattice structure. The increase in atomic mass (due to the replacement of Ba$^{2+}$ by Eu$^{2+}$) along with a decrease in the lattice constants (and unit cell volume as shown in Figure S2 in supplemental material) with increasing Eu$^{2+}$ should result in a reduced atomic displacement thus causing a decrease in the phonon anharmonicities. A detailed analysis of the quasiharmonic and intrinsic anharmonic contributions for all the phonon modes can be found in the supplemental material (Figure S8) that further elucidates that modes involving vibrations along the c-axis are more anharmonic in nature. The temperature ($T_C$) above which the mode P6 disappears is the crystallographic (tetragonal to cubic) transition temperature of BaTiO$_3$, also associated with the ferroelectric to paraelectric phase transition. A comparison of the temperature-dependent Raman spectra of Eu$_x$Ba$_{1-x}$TiO$_3$ (x= 0, 0.1, 0.2, 0.3, 0.4, and 1) from Figure 2, and Figures S5 and S6 in supplemental material can help determine this $T_C$. As shown in Figure 4(a), the $T_C$ decreases with increasing Eu$^{2+}$ in BaTiO$_3$ [35], a behaviour that is strikingly similar to the change in tetragonality (ratio of '*c*' to '*a*' lattice parameters) of the unit cell of BaTiO$_3$ with increasing Eu$^{2+}$ as shown in Figure 4(c). In fact, this similarity suggests that the reduction of ferroelectric order is clearly related to the suppression of phonon anharmonicity. Furthermore, the dielectric constant of Eu$_x$Ba$_{1-x}$TiO$_3$ as a function of temperature, as shown in Figure S9 in supplemental material, also reveals a decrease in the T$_C$ with increasing Eu$^{2+}$ thus corroborating our results shown in Figure 4. The decrease in phonon anharmonicity and the ferroelectric transition temperature (T$_C$) with Eu$^{2+}$-doping indicate that these parameters are coupled thus opening a new route to tailor the functionalities of ferroelectric materials [3,4].

**Conclusion:**

We have shown that phonon anharmonicities have an important association with ferroelectric ordering and the ferroelectric transition temperature in Eu$^{2+}$ doped BaTiO$_3$. Our experiments suggest that the competition between quantum fluctuations and ferroelectric ordering in Eu$_x$Ba$_{1-x}$TiO$_3$ systems maintains a delicate balance with phonon anharmonicities, suppression of which causes reduction of ferroelectric ordering. Tuning the phonon anharmonicity can therefore emerge as an additional route to tailor material functionalities that are coupled with the crystal structure.

**Methods:**

Polycrystalline BaTiO$_3$ and Eu$_x$Ba$_{1-x}$TiO$_3$ (x = 0.1, 0.2, 0.3, 0.4, and 1) powder samples were synthesized using solid state reaction route where BaCO$_3$, TiO$_2$, and Eu$_2$O$_3$ at stoichiometric ratio were used as precursors [36]. The well mixed and ground powder for BaTiO$_3$ was calcined at 1200 ˚C for 24 hrs in air. However, the powder for Eu$_x$Ba$_{1-x}$TiO$_3$ was calcined in reduced atmosphere (95% Ar-5% $H_2$) in order to reduce $Eu^{3+}$ to $Eu^{2+}$. Two more cycles of grinding and heating at 1200 ˚C were done followed by which the powders were pelletized and sintered at 1300 ˚C for 24 hrs in the same atmosphere. The polycrystalline samples were characterized by x-ray diffraction at room temperature and found to be phase pure. Further, the x-ray diffraction patterns were measured using PANalytical Empyrean x-ray diffractometer (Cu $K_\alpha$ line) as a function of temperature (Anton Paar TTK 450) from 90 to 450 K in order to probe the various structural phases and the corresponding lattice constants in $BaTiO_3$ and $Eu_{0.2}Ba_{0.8}TiO_3$. We would like to emphasize that based on our temperature-dependent x-ray diffraction measurements we have not been able to detect the presence of any impurity phase (of $Eu^{3+}$) within our experimental limit (see Section 1 in supplemental material). However, to be noted that the low temperature sample stage (referred as LTStage) gives rise to additional reflection peaks in the range of 40˚- 55˚ and near 75˚ which partially affect some of the $Eu_xBa_{1-x}TiO_3$ reflection peaks. The room temperature (300 K) data have been recorded on a different sample stage (referred as RTStage) which is free from those additional peaks and thus a comparison with the 300 K data allows us to identify the stage related artefacts. The additional peaks from the LTStage have been partly removed for clarity. A detailed comparison of the LTStage x-ray diffraction patterns with those of $Eu_xBa_{1-x}TiO_3$ are given in the supplemental material (Section S1, Figure S1). Further, magnetic measurements [36, 37] of the powder samples of $Eu_xBa_{1-x}TiO_3$ suggest an almost pure $Eu^{2+}$ state without any discernible impurity phase containing $Eu^{3+}$ (see Figure S3 and the discussion in supplemental material) thus confirming that the samples under study are in highly pure phase.

Raman spectra of $BaTiO_3$ and $Eu_xBa_{1-x}TiO_3$ were recorded in the backscattering geometry using a LabRAM HR Evolution Raman spectrometer equipped with a Peltier cooled charge coupled device (CCD) detector. The samples were mounted on a liquid nitrogen cooled Linkam heating stage and excited with the 514.5 nm line of an $Ar^+$-ion laser with a typical power of ~ 5 mW on the sample. Further tests were also carried out using the 532 nm laser line of a frequency doubles Nd-YAG laser and the results were found to be similar.

**Author contribution:**

B.P., S.Sar., and S.S. performed the Raman measurements and analyses. K.R. synthesized the samples under study and performed the dielectric and magnetic measurements. B.P performed the x-ray diffraction measurements. R.M., T.V., S.S. planned and supervised the project. B.P., S.Sar., T.V. and S.S. wrote the manuscript with contributions from all the authors.

**Acknowledgement:**

B.P. acknowledges the UGC, India, for fellowship. S.Sar would like to acknowledge the NGS fellowship. R.M. acknowledges Ministry of Education, Singapore (Grant no. MOE2015-T2-2-147). T.V. acknowledges support from the National Research Foundation under Competitive Research Program (NRF2015NRF-CRP001-015). S.S. acknowledges SERB, India, for research grant (Project no. SERB/ECR/2016/001376) and financial supports from DST-FIST (Project No. SR/FST/PSI-195/2014(C)). Authors acknowledge Mr. Manoj Prajapat and Ms. Anjali Rathore at IISER Bhopal for their help during XRD measurements.

Table-1: Assignment of the Raman active phonon modes (and their frequencies in cm$^{-1}$ unit) of BaTiO$_3$ and Eu$_{0.2}$Ba$_{0.8}$TiO$_3$ at 78 K (Rhombohedral), 230 K (Orthorhombic), and 300 K (Tetragonal) obtained by fitting the experimental data.

| Mode Name | Mode Symmetry | BaTiO$_3$ | | | Eu$_{0.2}$Ba$_{0.8}$TiO$_3$ | | |
|---|---|---|---|---|---|---|---|
| | | At 78 K | At 230 K | At 300 K | At 78 K | At 230 K | At 300 K |
| P1 | $A_1(TO_1)$ | 168 | 155 | Absent | 161 | 154 | Absent |
| P2 | $A_1(LO_1)$ | 188 | 195 | Absent | 186 | 192 | Absent |
| P3 | Disorder induced Raman modes* | 222 | 218 | 218 | 220 | 209 | 208 |
| P4 | | 240 | Absent | Absent | 240 | Absent | Absent |
| P5 | $A_1(TO_2)$ | 259 | 258 | 264 | 255 | 258 | 253 |
| P6 | $B_1$ | 311 | 309 | 308 | 310 | 307 | 306 |
| P7 | $A_1(LO_2)/E(LO_3)$ | 489 | Absent | Absent | 490 | Absent | Absent |
| P8 | $E(TO_4)$ | 531 | 524 | 520 | 528 | 516 | 514 |
| P9 | $A_1(TO_3)$ | Absent | 551 | 556 | Absent | 555 | 558 |
| P10 | $A_1(LO_3)/E(LO_4)$ | 716 | 717 | 719 | 716 | 718 | 720 |

*The modes P3 and P4 are assigned to disorder induced Raman active modes [18].

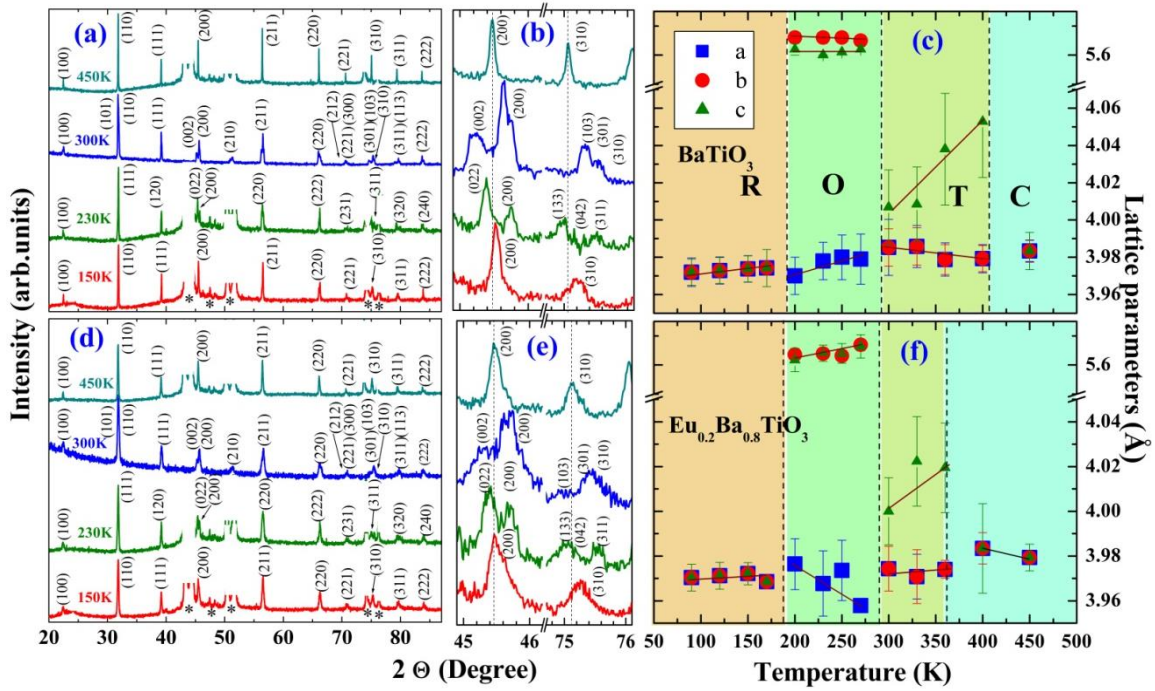

Figure 1: X-ray diffraction patterns of BaTiO$_3$ (a, b) and Eu$_{0.2}$Ba$_{0.8}$TiO$_3$ (d, e) at various temperatures with corresponding (hkl) values. The middle column (b, e) shows the specific reflection peaks that correspond to the structural phase transitions. The right column shows the lattice parameters of (c) BaTiO$_3$ and (f) Eu$_{0.2}$Ba$_{0.8}$T
iO$_3$ as a function of temperature as it undergoes the structural phase transitions (from R=Rhombohedral to O=Orthorhombic to T = Tetragonal to C = Cubic phases shown by the shaded regions). The solid lines in (c) and (f) are guide to the eye. The additional peaks marked with asterisk (*) symbols in the regions between 40-55° and near 75° in (a) & (d) are due to the low temperature sample stage (LTStage) which have been partly removed from the data for clarity. The room temperature (300 K) data do not have these additional peaks and hence the (210) sample peak is clearly visible.

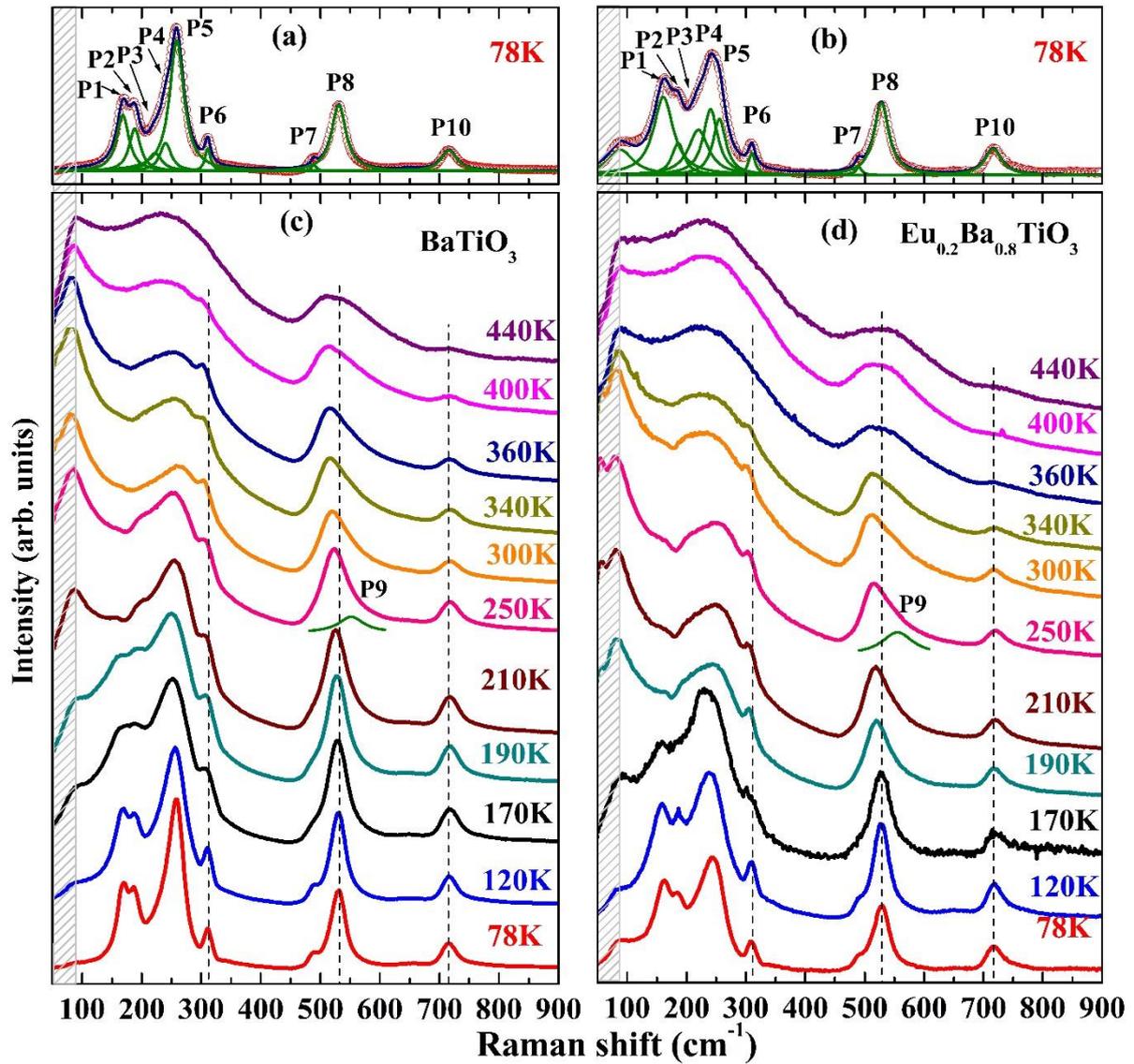

Figure 2: Raman spectra of $BaTiO_3$ and $Eu_{0.2}Ba_{0.8}TiO_3$ at a few typical temperatures showing the evolution of the phonon modes with temperature. The shading indicates the region affected by the optical band-pass filter as discussed in the text. The spectra have been analyzed using Lorentz functions to identify the phonon modes P1 to P10 (shown in green) in (a) and (b). The mode P9 (shown in green in c and d) is present above ~ 190 K in the orthorhombic and tetragonal phases.

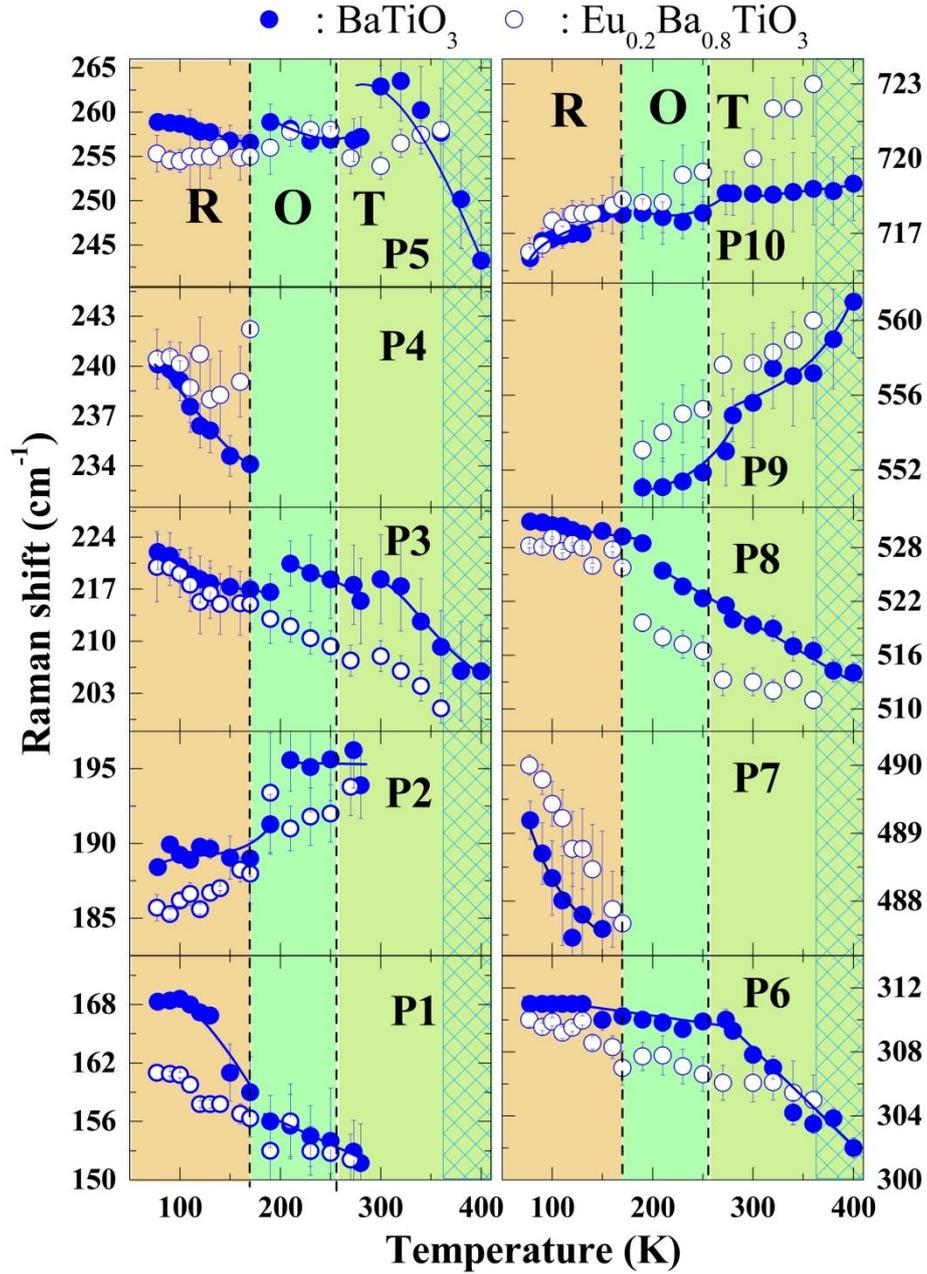

Figure 3: Frequency of all the phonon modes (P1 to P10) as a function of temperature. All the modes respond to the structural phase transitions as the temperature varies. Effect of $Ba^{2+}$ replacement by $Eu^{2+}$ is also evident in all the modes manifested as the change in phonon anharmonicity, discussed in the text. The colour-shaded regions indicate the various structural phases as indicated by the letters R: Rhombohedral, O: Orthorhombic, and T: Tetragonal. The cross-shaded region indicates the Cubic phase for $Eu_{0.2}Ba_{0.8}TiO_3$. The solid lines are guide to the eye.

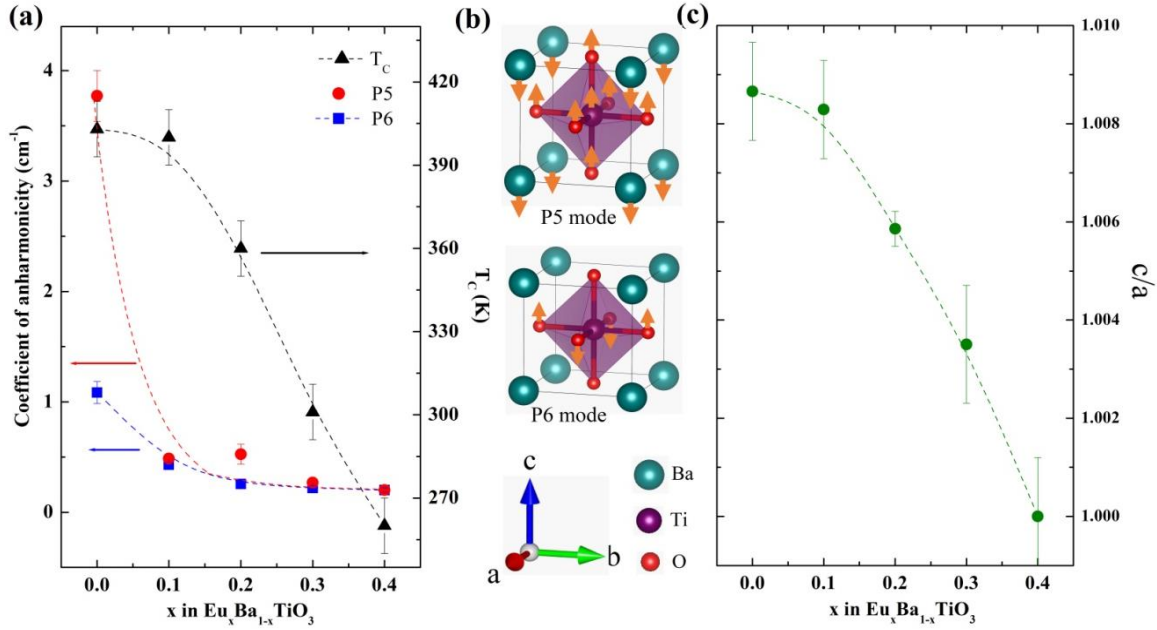

Figure 4: (a) Dependence of the coefficient of phonon anharmonicity (as described in eqn. 2) on the doping concentration of $Eu^{2+}$ by replacing $Ba^{2+}$ for the modes at 265 cm$^{-1}$ (P5) and 308 cm$^{-1}$ (P6). A decrease in phonon anharmonicity is evident with increasing $Eu^{2+}$. Similarly, the para-to-ferroelectric phase transition temperature ($T_C$) also decreases with increasing $Eu^{2+}$, as discussed in the main text. (b) Schematic of the atomic vibrations for the phonon modes at 265 cm$^{-1}$ (P5) and 308 cm$^{-1}$ (P6). (c) Reduction in tetragonality (ratio of '$c$' to '$a$' lattice parameters) of $Eu_xBa_{1-x}TiO_3$ at room temperature with increasing $Eu^{2+}$. The solid symbols (in a and c) represent the experimental data while the dashed lines are guide to eye.

# Effect of Phonon Anharmonicity on Ferroelectricity in $Eu_xBa_{1-x}TiO_3$

Bommareddy Poojitha,[1,†] Km Rubi,[2,†] Soumya Sarkar,[3,4,†] R. Mahendiran,[2,3,*]
T. Venkatesan,[2,3,4,*] and Surajit Saha,[1,*]

[1]Department of Physics, Indian Institute of Science Education and Research, Bhopal 462066, INDIA

[2]Department of Physics, National University of Singapore, Singapore 117551, SINGAPORE

[3]NUSNNI-NanoCore, National University of Singapore, Singapore 117411, SINGAPORE

[4]NUS Graduate School for Integrative Sciences and Engineering, Singapore 117456, SINGAPORE

† *Equal contributors*

*Correspondence: S. Saha- email: surajit@iiserb.ac.in (+91-755-2691225); T. V. - email: venky@nus.edu.sg (+65-65165187); R. M.- email: phyrm@nus.edu.sg (+65-65162616)

This supplemental material file contains additional data on x-ray diffraction, saturation magnetization, Raman spectroscopic, and dielectric constant measurements of $Eu_xBa_{1-x}TiO_3$. A detailed description on the Raman modes of $BaTiO_3$ is also provided. The figures and respective details are given below.

## S1. Phase purity of $Eu_xBa_{1-x}TiO_3$

*X-ray diffraction:* To ascertain the phase purity of the powder samples of $Eu_xBa_{1-x}TiO_3$, we have analysed the low temperature x-ray diffraction data carefully and observed no discernible presence of impurity phase(s) containing $Eu^{3+}$. The temperature-dependent XRD data show very strong additional reflections arising from the low-temperature sample stage (later referred as LTStage) which were removed while analysing the data. To rule out the possibilities, we have included the XRD data of the LTStage at a few temperatures and compared with the XRD data of $Eu_xBa_{1-x}TiO_3$ at the same temperatures thus distinguishing and confirming that the additional reflection peaks arise from the LTStage (environment) and not from impurity.

In the Figure S1, we present the x-ray diffraction patterns of the LTStage in black colour and also the $Eu_xBa_{1-x}TiO_3$ (x=0 in red and x= 0.2 in blue colour) mounted on the LTStage at 300, 230, and 150 K. The experimental data at 300 K are also compared with the theoretically calculated diffraction pattern (using VESTA software) for the tetragonal phase of $BaTiO_3$. A comparison of the diffraction patterns clearly indicates that the strong reflections near 40-55° and near 75° (which were removed for clarity and to analyse the data) are due to the LTStage and not due to any impurity phase. Moreover, all the expected reflection peaks in the tetragonal phase are observed in the experimental data. Hence, we would like to emphasize that our prepared samples of $Eu_xBa_{1-x}TiO_3$ do not contain any discernible impurity. However, it is important to note that there are two weak reflection peaks near 47.5° and 48.5° that are present in the patterns of $Eu_xBa_{1-x}TiO_3$ (x=0 and 0.2), as shown in Figure S1(a) and in the enlarged Figure S1(b). It needs to be mentioned that for regular XRD measurements at our centre (i.e. when the sample is kept at room temperature and the sample temperature is not varied), a separate sample stage (here referred as RTStage) is used which does not have any unwanted diffraction peak at these angles (rather at any angle). Hence, our $Eu_xBa_{1-x}TiO_3$ samples when mounted on the RTStage, the two weak reflection peaks near 47.5° and 48.5° are not observed in the diffraction data, as shown in Figure S1(b) (diffraction data labelled as EBT without LTStage). Therefore, this comparison clarifies that these two weak peaks are again not due to any impurity phase.

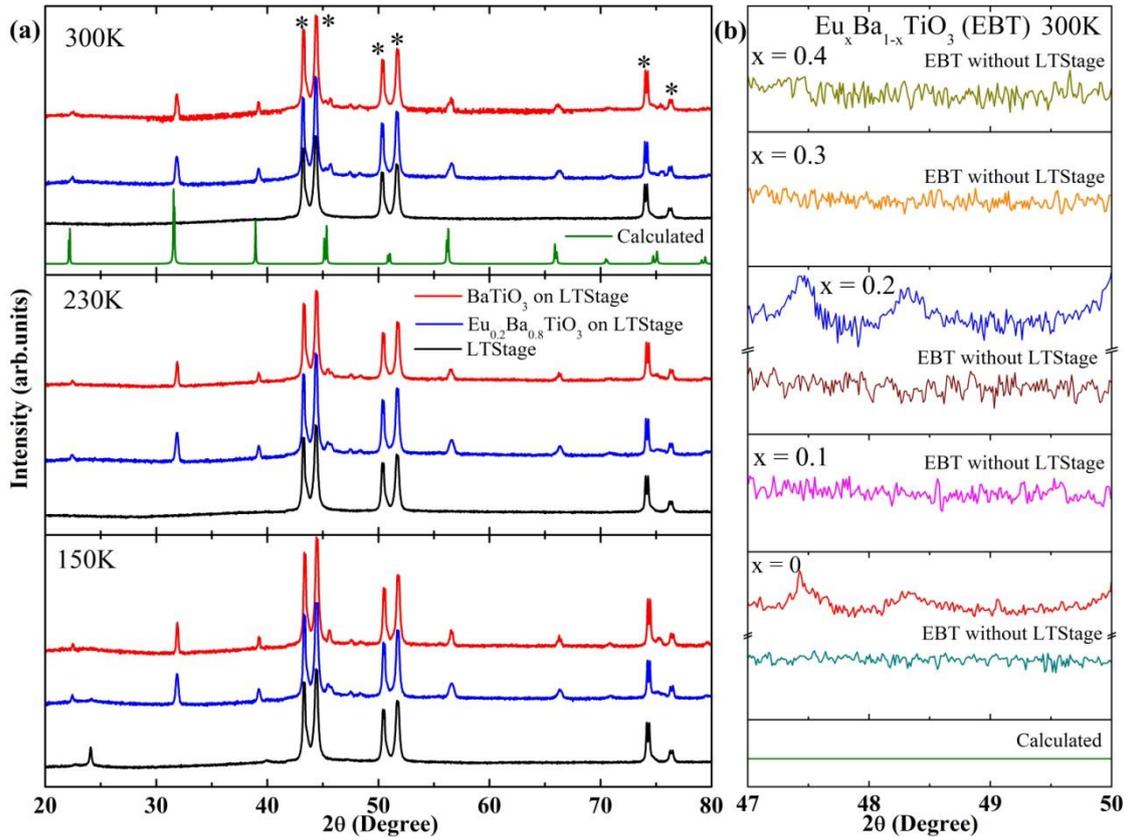

Figure S1: (a) X-ray diffraction pattern of $Eu_xBa_{1-x}TiO_3$ (mounted on the low temperature sample stage i.e. LTStage) at 300, 230, and 150 K. The LTStage has very strong reflections in the range of 40-55° and near 75° (data shown in black colour). These strong reflections are also seen in the diffraction patterns of $Eu_xBa_{1-x}TiO_3$ (red colour for x=0 and blue for x=0.2) at all temperatures. In addition, two weak reflection peaks are observed at 47.5° and 48.5° which are shown in enlarged figure (b). The two weak peaks are not observed in the $Eu_xBa_{1-x}TiO_3$ samples when measured without the LTStage, as discussed above.

In order to have a clarity on the possible presence of $Eu^{3+}$ impurity phase, we have further analysed the XRD data. If Eu-ion is present in 3+ state, it will introduce charge imbalance. In order to self-compensate the charge, $Eu^{3+}$ replaces both the $Ba^{2+}$ and $Ti^{4+}$ thus leading to an increase in the unit cell volume [1]. Interestingly, we do not observe any such increase in unit cell volume, rather, we observe a decrease in the volume with increasing Eu-ion content. We have shown the dependence of the volume and the lattice parameters at room temperature as a function of Eu-content in Figure S2. It, therefore, suggests that if at all $Eu^{3+}$ is present, it is insignificant in proportion.

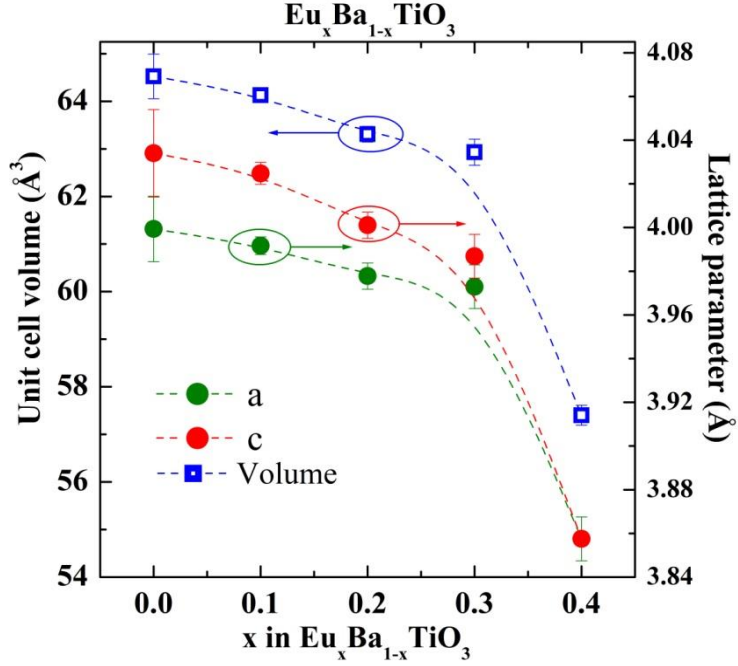

Figure S2: Unit cell volume and lattice parameters of $Eu_xBa_{1-x}TiO_3$ as a function of x at room temperature.

Further, Raman spectroscopy is known to be a very sensitive technique to probe impurity phase(s) present in samples. In Eu-doped $BaTiO_3$, if Eu-ions exist in 3+ state it leads to a deformation of the $TiO_6$ octahedron thus giving rise to a phonon band near 830 cm$^{-1}$ [1]. Importantly, our Raman spectra (Figure 2 in the main text and Figures S4-S6) do not show the presence of any band near 830 cm$^{-1}$ thus providing an additional evidence of the absence of $Eu^{3+}$ ions in our samples.

Above all, to further clarify on the possible presence of $Eu^{3+}$ impurity phase, magnetic measurements were also performed. The magnetization data on our $Eu_xBa_{1-x}TiO_3$ samples have been reported earlier [2,3] by the co-authors. The near saturation magnetization of $Eu_xBa_{1-x}TiO_3$ at a magnetic field of 5T as a function of x is shown in Figure S3. The magnetization data [2,3] indicate that the effective magnetic moment for Eu-ion in the $Eu_xBa_{1-x}TiO_3$ samples is ~ 7.8 $\mu_B$ which is very close to the theoretical value of 7.94 $\mu_B$ for $Eu^{2+}$ and this experimental estimation would further approach the theoretical value at even higher magnetic fields. The magnetic moment of $Eu^{3+}$ is ~ 3.45 $\mu_B$ as predicted by Van Vleck [4,5] and, therefore, presence of any significant proportion of $Eu^{3+}$ would reduce the effective magnetic moment (of Eu-ion) drastically. It needs to be brought to the attention that several reports [6-8] on nearly pure $Eu^{2+}$-based compounds have measured an effective magnetic moment of ~7 $\mu_B$ which is even lower than what we have observed. Nonetheless, being highly conservative one may still assume that there is a small fraction of $Eu^{3+}$ present in our samples and estimate an approximate fraction by using the relation $\mu_{eff}^2 = Z\mu_{Eu^{2+}}^2 + (1-Z)\mu_{Eu^{3+}}^2$; where $\mu_{eff}$ is the effective magnetic moment while $\mu_{Eu^{2+}}$ and $\mu_{Eu^{3+}}$ are the magnetic moments of $Eu^{2+}$ and $Eu^{3+}$ ions, respectively, and $Z$ gives the percentage of the $Eu^{2+}$ ions present. Thus, based on the experimental value of the magnetic moment a conservative approach would confirm the presence of more than 98% of Eu-ions in the 2+ state in our samples which would asymptotically approach the value of 100% if the magnetic measurements could possibly be performed at much higher magnetic fields.

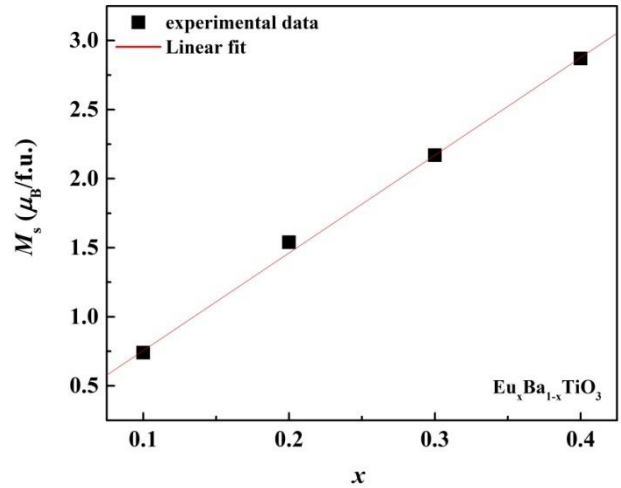

Figure S3: Saturation magnetization ($M_S$) of $Eu_xBa_{1-x}TiO_3$ as a function of x at a magnetic field of 5 T which confirms that the Eu-ions are in 2+ state. An increase in Eu-content linearly increases the $M_S$ thus confirming that the Eu-ions remain in the 2+ state in all the $Eu_xBa_{1-x}TiO_3$ samples.

To summarize, presence of any small amount of impurity phase containing $Eu^{3+}$ could not be detected within the measurement limits of the techniques used (XRD, Raman, and magnetization measurements). Therefore, we believe that the impurity phase, if any, will have insignificant effect on the observations and arguments made in the manuscript.

## S2. Description of the Raman modes of BaTiO₃

$BaTiO_3$, in both the ferro- and para-electric phases it has one formula unit (i.e., 5 atoms) per unit cell that gives rise to twelve optical phonon modes. In the paraelectric-cubic phase none of the phonons ($3F_{1u}+F_{2u}$) are Raman active whereas in the ferroelectric-tetragonal phase there are $3A_1$, $B_1$, and $4E$ Raman modes. The soft phonon mode $F_{1u}$ in the cubic phase splits in to two modes with symmetries $A_1$ (appears at around 265 cm$^{-1}$) and *E* (appears at around 35 cm$^{-1}$) in the tetragonal phase. One of the very early reports was by Perry and Hall [9] that studied the effect of thermally-induced structural phase transitions on the Raman active phonons emphasizing the hysteretic behaviour of the $A_1$(TO$_2$) soft phonon mode at ~ 265 cm$^{-1}$ (O-Ti-O bending vibration) and its abrupt changes in frequencies due to the structural transitions to orthorhombic and rhombohedral phases. The other two $A_1$ phonon modes in BaTiO$_3$ appear at around 520 and 720 cm$^{-1}$ that arise due to the transverse and longitudinal Ti-O stretching vibrations of TiO$_6$ octahedra. A unique feature of BaTiO$_3$ is the depolarized '*dip*' at around 180 cm$^{-1}$ with an asymmetric line-shape of the 265 cm$^{-1}$ ($A_1$) phonon mode which has been attributed to an interference effect associated with the anharmonic coupling of its three $A_1$ phonons [10-12]. The '*dip*', however, disappears in nano-particles [13,14] and in some cases in polycrystalline powder [13,15] possibly indicating no-coupling of the $A_1$ phonons. The '*dip*' at 180 cm$^{-1}$ appears as a '*peak*' under polarization has been assigned to $A_1$(TO) as well as *E*(TO) [16-18]. The overdamped *E*(TO) soft mode appears at around 35 cm$^{-1}$ has a very broad linewidth (~100 cm$^{-1}$) [11,16] while the other *E* modes appear as weak peaks at around 466 cm$^{-1}$ (LO) and 489 cm$^{-1}$ (TO) [11,16,17]. The mode at 308 cm$^{-1}$, which is reported to be seen in the low temperature crystallographic (ferroelectric) phases [9,16], is assigned as a mixed $B_1$+*E* mode originating from the Ti-O$_3$ torsional vibration and is a result of the splitting of the 'silent' $F_{2u}$ mode in paraelectric-cubic phase [11,16,18,19]. Notably, the mode assignments in literature are normally done in the tetragonal phase of BaTiO$_3$ because the complex domain structures in orthorhombic and rhombohedral phases cause a complete depolarization of the Raman spectra in these respective phases [16,18].

**S3. Raman spectra of Eu$_x$Ba$_{1-x}$TiO$_3$ (x = 0, 0.1, 0.2, 0.3, 0.4, and 1) at room temperature and low temperatures**

Figure S4 below shows the Raman spectra of Eu-doped BaTiO$_3$ at room temperture with varying Eu concentration. A clear evolution of the spectra is obsevred with increasing doping of Eu. One of the important phonon modes is at 308 cm$^{-1}$ ($A_1$), marked by the dotted line, which is a representative of the low temperature (ferroelectric) phase of BaTiO$_3$ because this mode is present in the ferroelectric low temperature structural phases while it disappears in the paraelectric-cubic phase [9-11,19,20]. It, therefore, can be suggested based on Figure S4 that Eu$_x$Ba$_{1-x}$TiO$_3$ is a room temperature ferroelectric only for x = 0 to 0.3. For x=0.4 and above, it is no longer room temperature ferroelectric. Figures S5 and S6 show the Raman spectra of Eu$_x$Ba$_{1-x}$TiO$_3$ for x=0.1, 0.3, 0.4, and 1 at a few temperatures. The mode near 308 cm$^{-1}$ becomes very weak at 400 K for Eu$_{0.1}$Ba$_{0.9}$TiO$_3$ and 315 K for Eu$_{0.3}$Ba$_{0.7}$TiO$_3$ above which they were not observed while in Eu$_{0.4}$Ba$_{0.6}$TiO$_3$ it is weak at all the low temperatures and its presence can barely be seen at temperatures between 240-280 K. This mode is not expected in pure

EuTiO$_3$ and hence is absent. The presence of the 308 cm$^{-1}$ mode until a given maximum temperature signifies the tetragonal/ferroelectric phase and the ferroelectric-paraelectric transition temperature.

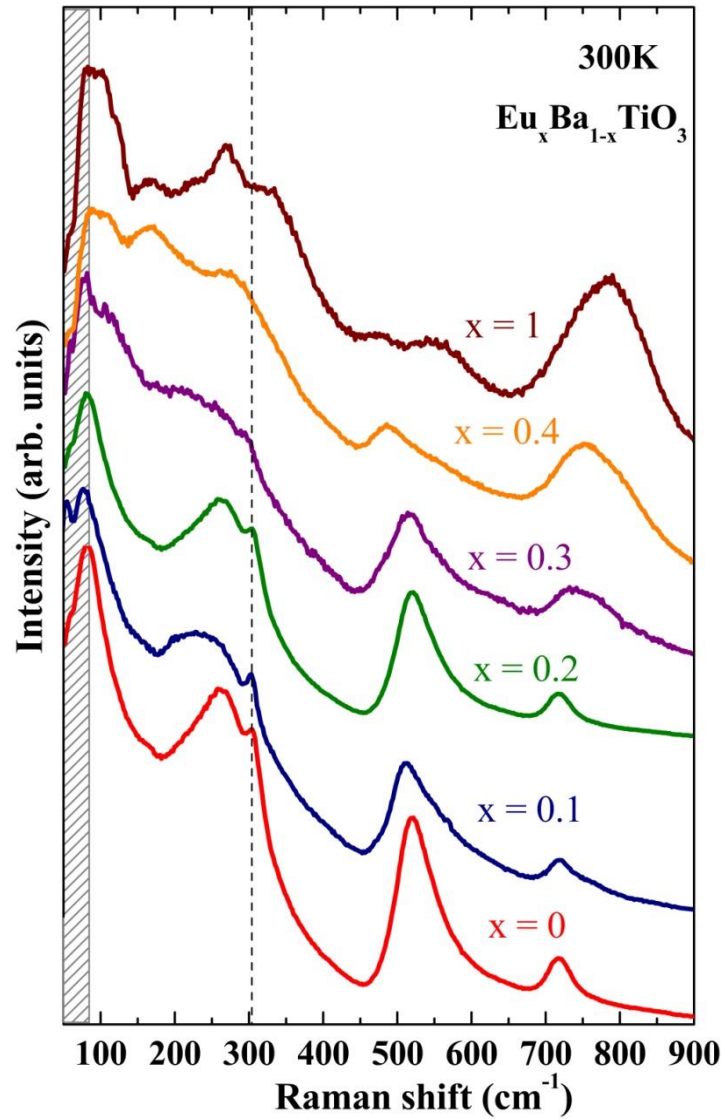

Figure S4: Raman spectra of Eu$_x$Ba$_{1-x}$TiO$_3$ with x = 0, 0.1, 0.2, 0.3, 0.4, and 1 recorded at 300 K. As evident from the data, the phonon bands evolve with increasing content of Eu as it goes from pristine BaTiO$_3$ to pristine EuTiO$_3$. The phonon modes are similar for low value of x (i.e. x = 0 to 0.2) while for x = 0.3 and above, signatures of EuTiO$_3$ start to appear in the spectra. The mode at 308 cm$^{-1}$ vanishes for x = 0.4 and above in Eu$_x$Ba$_{1-x}$TiO$_3$. The shading indicates the region that is affected by the optical band-pass filter and is not considered for data analysis.

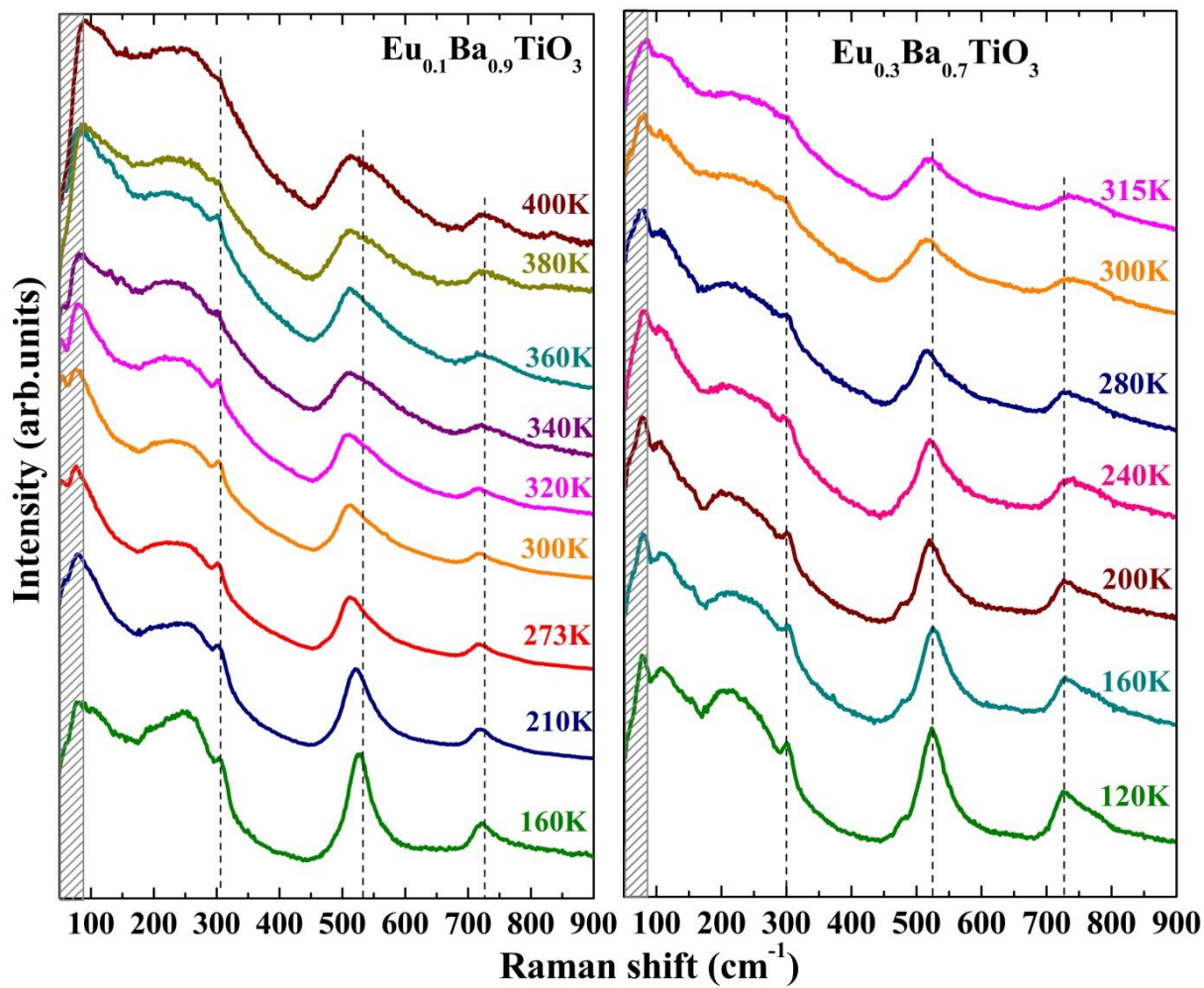

Figure S5: Raman spectra of (i) $Eu_{0.1}Ba_{0.9}TiO_3$ and (ii) $Eu_{0.3}Ba_{0.7}TiO_3$ at a few typical temperatures showing the evolution of the phonon modes with varying temperatures. The mode near 308 cm$^{-1}$ (identified by dashed lines) becomes very weak at 400 K for $Eu_{0.1}Ba_{0.9}TiO_3$ and 315 K for $Eu_{0.3}Ba_{0.7}TiO_3$ above which they were not observed. Presence of the 308 cm$^{-1}$ mode until a given temperature signifies the tetragonal/ferroelectric phase. The shading indicates the region that is affected by the optical band-pass filter.

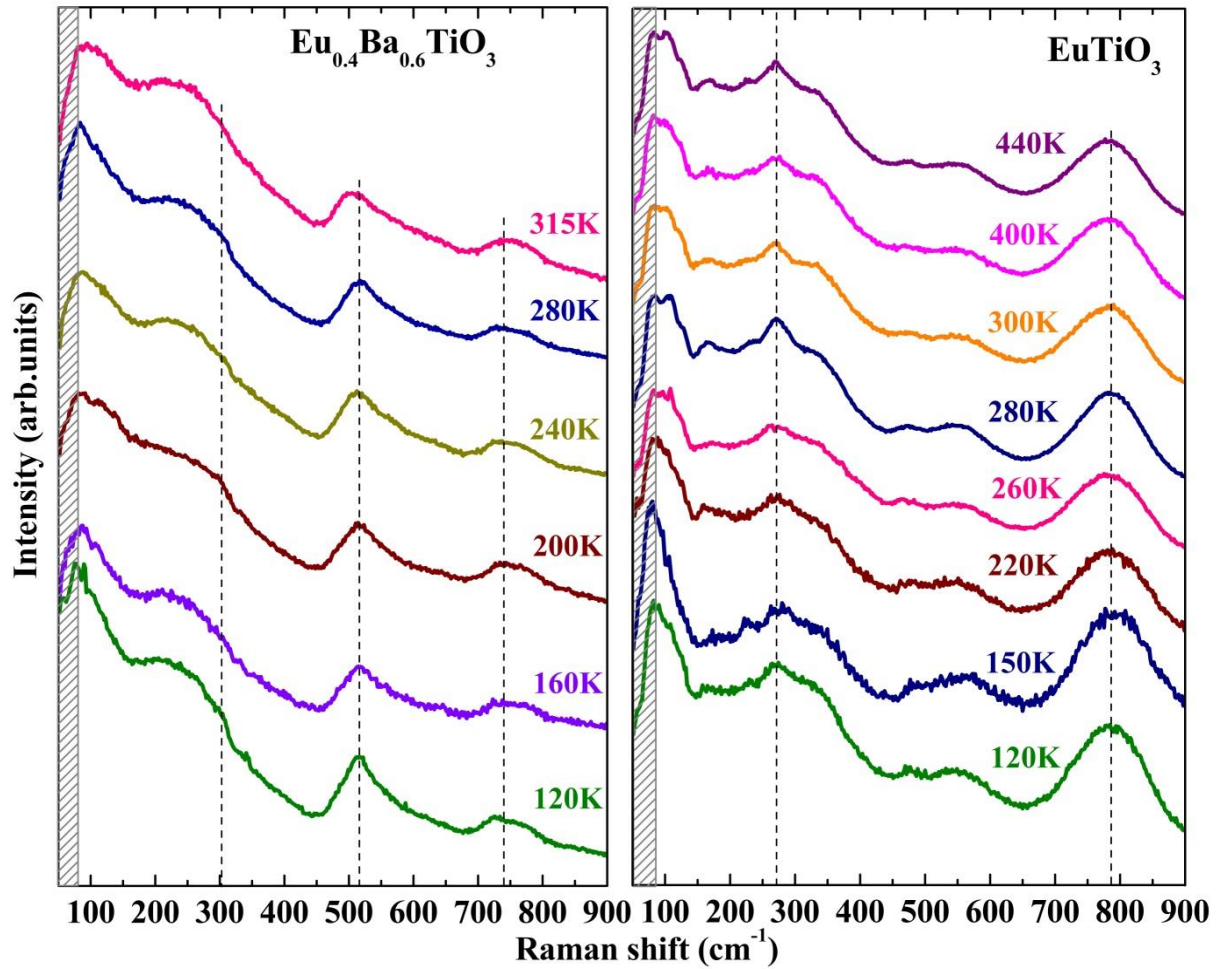

Figure S6: Raman spectra of (i) $Eu_{0.4}Ba_{0.6}TiO_3$ and (ii) $EuTiO_3$ at a few typical temperatures showing the evolution of the phonon modes with varying temperatures. As evident from the spectra, the spectrum of $Eu_{0.4}Ba_{0.6}TiO_3$ is quite different from the pure $BaTiO_3$ spectrum where influence of $EuTiO_3$ is clearly present. Nonetheless, the mode near 308 cm$^{-1}$ is weak in $Eu_{0.4}Ba_{0.6}TiO_3$ at all the low temperatures and its presence can barely be seen at temperatures between 240-280 K. This mode is not expected in pure $EuTiO_3$ and hence is absent. The shaded region of the spectrum that is affected by the optical band-pass filter and hence not considered for analysis.

## S4. Phonon anharmonicity of the 265 and 308 cm$^{-1}$ modes

The modes P5 (265 cm$^{-1}$) and P6 (308 cm$^{-1}$) have been fitted in Figure S7 with cubic anharmonic equation given in eqn. 2 in main text in order to quantify the strength of anharmonicity for these modes. The derived quantity is then compared as a function of Eu doping in BaTiO$_3$ (shown in Figure 4 in the main text).

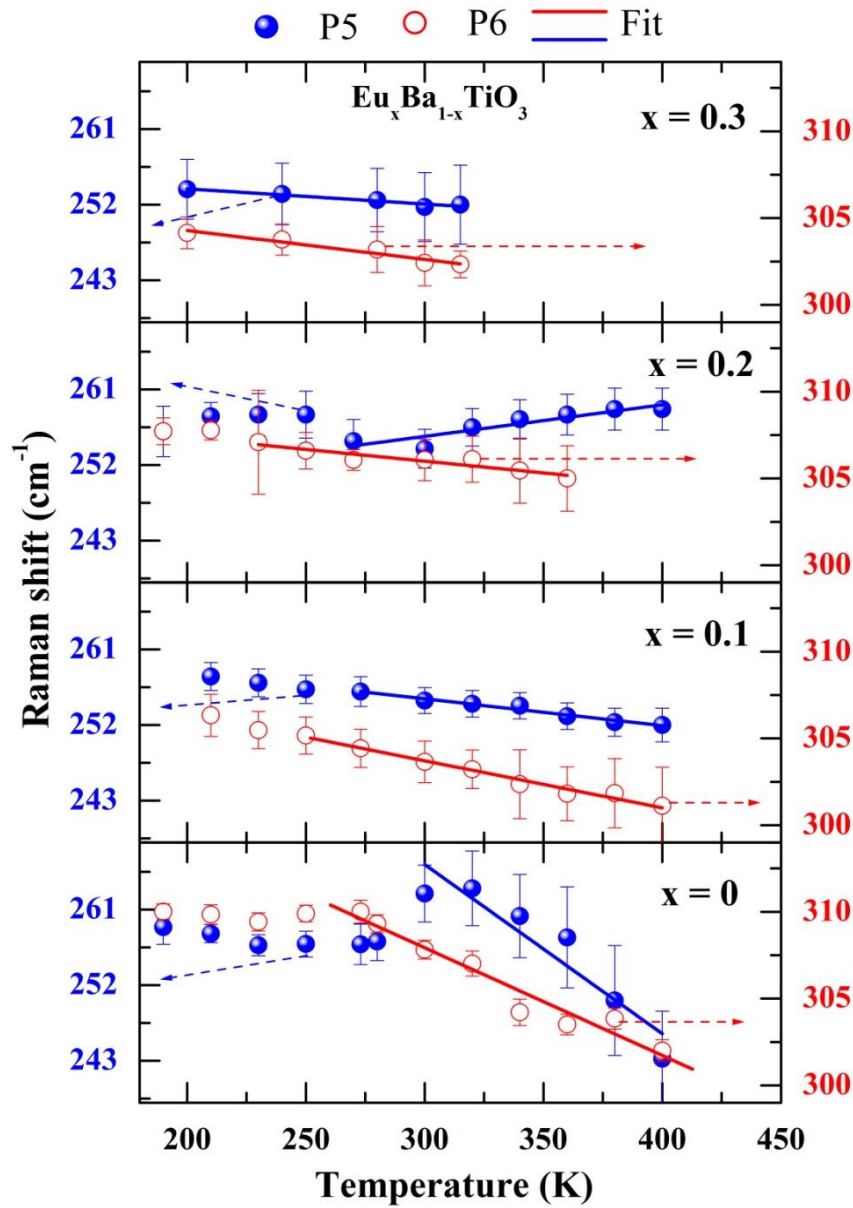

Figure S7: The temperature dependence of the P5 (265 cm$^{-1}$) and P6 (308 cm$^{-1}$) phonon modes for Eu$_x$Ba$_{1-x}$TiO$_3$ (x = 0, 0.1, 0.2, 0.3). The phonon frequency has been fitted (solid lines) with cubic anharmonic equation (eqn. 2 in main text) in order to derive the strength of anharmonicity with varying x.

**S5. Phonon intrinsic anharmonicity**

As described in the eqn. 1 in main text, in absence of spin-phonon and electron-phonon coupling, the temperature dependence of frequency of a phonon ($i$) may be expressed as [20,21]:

$$\omega^i(T) = \omega^i(0) + \Delta\omega^i_{qh}(T) + \Delta\omega^i_{anh}(T)$$

where, $\omega^i(0)$ is the phonon frequency at 0 K. The term $\Delta\omega^i_{qh}(T)$ is the change in frequency due to the quasiharmonic contribution arising from the change in lattice volume (i.e., force constant) without changing the phonon population. The term $\Delta\omega^i_{anh}(T)$ is the change in frequency due to the intrinsic anharmonic contribution arising from the real part of the self-energy of a phonon decaying into two (cubic anharmonicity) or three (quartic anharmonicity) phonons. The quasiharmonic contribution ($\Delta\omega^i_{qh}(T)$) to the phonon frequency depends purely on the change in lattice volume which is related to the ($i^{th}$) mode Grüneisen parameter as $\gamma^i = -\frac{dln\omega^i}{dlnV} = \frac{B_0 d\omega^i}{\omega^i dP}$ [20,21] where $B_0$ is the bulk modulus and $V$ is the unit cell volume at a pressure $P$. Based on the pressure-dependent Raman studies of BaTiO$_3$ [12,15], we can consider the Grüneisen parameter $\gamma^i \sim 1$ for all the phonon modes in the tetragonal phase, except for the $A_1$ mode at 265 cm$^{-1}$ for which the $\gamma^i \sim -11$ because this mode undergoes an anomalous decrease in frequency with increasing pressure [15]. Importantly, in absence of high pressure Raman and x-ray diffraction studies of BaTiO$_3$ at low temperatures, we have assumed $\gamma^i \sim 1$ for all the phonons in the rhombohedral and orthorhombic phases. This may be justified because in common solids the Grüneisen parameter as $\gamma^i \sim 1$ [20,21]. Using the reported values of bulk modulus [22] and the above parameters, one can estimate the quasi harmonic contribution considering the relation [20,21], $\Delta\omega^i_{qh}(T) = \omega^i(T) - \omega^i(LT) = -\gamma^i\omega^i(LT)\frac{V(T)-V(LT)}{V(LT)}$, where $T$ stands for the variable sample temperature and $LT$ is the lowest temperature of measurement in the respective structural phases. In order to find the temperature-dependent unit cell volume, we refer to the x-ray diffraction patterns of the polycrystalline powder samples of BaTiO$_3$ and Eu$_{0.2}$Ba$_{0.9}$TiO$_3$ shown in Figure 1 and the corresponding lattice parameters (for all the structural phases) as a function of temperature that show a good agreement with the previous reports [23]. Therefore, the quasiharmonic change in frequency ($\Delta\omega^i_{qh}(T)$) is estimated using the above relation which is then subtracted from the experimental values of mode frequencies to obtain the change in frequency arising solely due to the intrinsic phonon anharmonic effects ($\Delta\omega^i_{anh}(T)$). Figure S8 shows the estimated $\frac{\Delta\omega^i_{anh}(T)}{\omega(LT)}$ % as a function of temperature for all the modes for both BaTiO$_3$ and Eu$_{0.2}$Ba$_{0.8}$TiO$_3$. The most interesting behaviour can be seen for the mode P5 in BaTiO$_3$ ($A_1$ mode at 265 cm$^{-1}$) which does not show any significant anharmonicity at the low temperature rhombohedral and orthorhombic phases ($\frac{\Delta\omega_{anh}^i(T)}{\omega(LT)} \sim 0\ \%$) but has the highest intrinsic anharmonicity of up to ~ 20 % (considering $\gamma^i \sim -11$, shown by red-coloured solid circles in Figure S8) in the tetragonal phase as compared to the other phonons of BaTiO$_3$. Even if we assume $\gamma^i \sim 1$ (shown by blue-coloured solid circles in Figure S8) for this mode in the tetragonal phase (like it is assumed for all the phonons in other phases), this mode still shows a high intrinsic anharmonicity of ~7%. In contrast, when Ba is partially replaced by Eu, the anharmonicity of this mode (in Eu$_{0.2}$Ba$_{0.8}$TiO$_3$) almost disappears in the tetragonal phase. To recall, this mode arises from the O-Ti-O bending motions of TiO$_6$ octahedra coupled with the Ba displacements where all the atoms vibrate along the c-axis [24]. A replacement of Ba$^{2+}$ by Eu$^{2+}$ increases the effective mass and decreases the unit cell volume which therefore should reduce the displacements of the participating atoms and hence the phonon anharmonicity. Earlier reports [10,12] had considered an anharmonic coupling of this mode with the other $A_1$ modes in the tetragonal phase of BaTiO$_3$. Here, we have quantified the anharmonic contribution to the mode as a function of temperature in all the low temperature phases. It may be noted that the modes P1, P2, P3, P4, P9, and P10 show significant anharmonicity which gets reduced/suppressed upon incorporating Eu$^{2+}$.

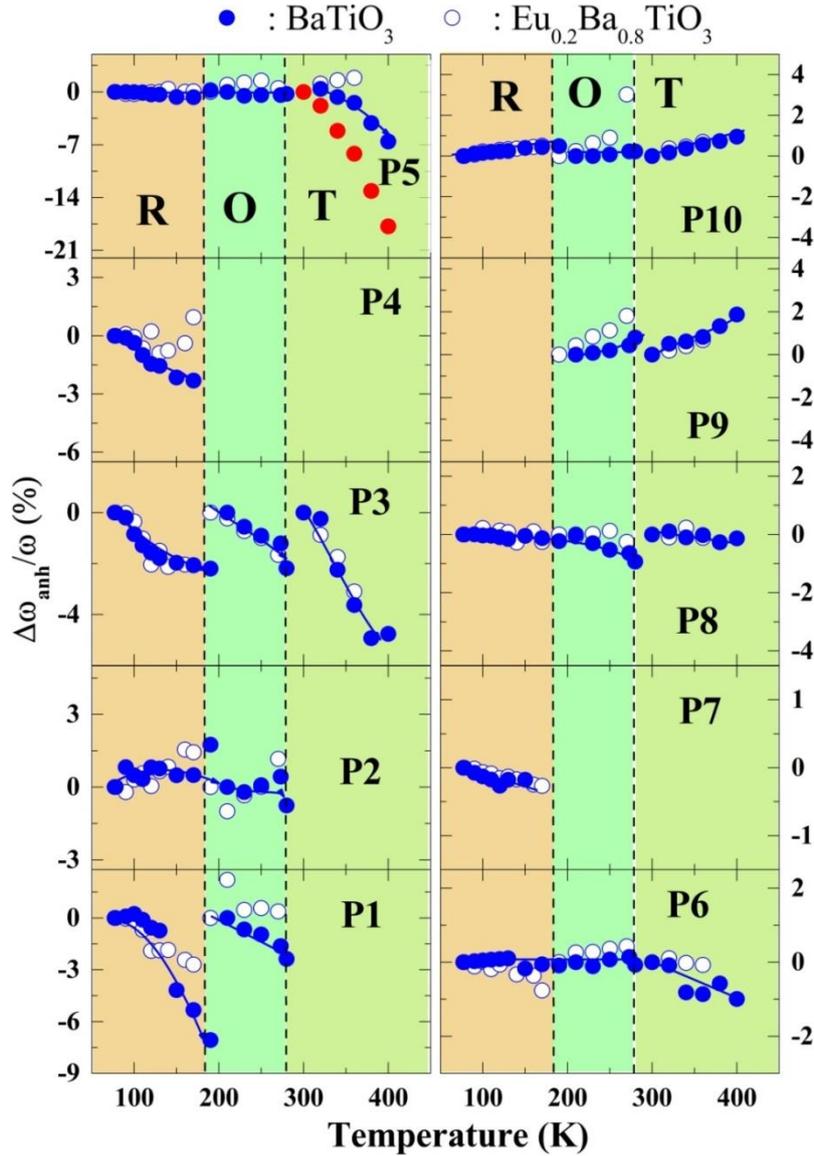

Figure S8: Estimated intrinsic phonon anharmonicity for all the modes (P1 to P10) of BaTiO$_3$ and Eu$_{0.2}$Ba$_{0.8}$TiO$_3$ as a function of temperature. Mode P5 shows a very high anharmonicity in tetragonal phase. The blue circles are estimated (as per the procedure discussed above) using Grüneisen parameter as $\gamma^i \sim 1$ while red circles are the estimates using as $\gamma^i \sim -11$. Eu incorporation suppresses the anharmonicities in most of the modes. The solid lines are guide to eye. The colour-shaded regions indicate the various structural phases - R: Rhombohedral, O: Orthorhombic, and T: Tetragonal, respectively.

## S6. Dielectric Measurement

The capacitance of the samples having parallel plate capacitor geometry was measured using Agilent 4294A impedance analyzer while cooling in a closed cycle refrigerator cryostat. From the measured capacitance, the dielectric constant ($\varepsilon$) was calculated using the relation, $\varepsilon = \frac{Ct}{\varepsilon_0 A}$, where $t$ is the thickness, $A$ is the cross section area of the sample and $\varepsilon_0$ is vacuum permittivity. In order to form a parallel plate capacitor, electrodes with silver paint were made on the largest surfaces of the sample. The samples were polished before making contacts in order to obtain a smooth surface and better sample-electrode interface.

Figure S9 shows the temperature dependence of the dielectric constant ($\varepsilon$) measured with the frequency of 1 MHz for samples $Eu_xBa_{1-x}TiO_3$ ($x$ = 0.0, 0.2 and 0.4). The undoped compound, i.e. pure $BaTiO_3$ exhibits a peak in $\varepsilon$ ($\varepsilon_{max}$ ~ 5700) at $T$ = 397 K which decreases as the $Eu^{2+}$ substitution for $Ba^{2+}$ site increases. The $\varepsilon(T)$ for $x$ = 0.0 display three sharp anomalies at the temperatures 397 K, 286 K and 194 K corresponding to paraelectric/ferroelectric or cubic to tetragonal (C-T), tetragonal to orthorhombic (T-O) and orthorhombic to rhombohedra (O-R) phase transitions, respectively. On the other hand, the doped compounds show only one broad and clear peak in $\varepsilon$ (marked by arrow) but at lower temperatures (at $T$ ~ 343 K and 254 K for $x$ = 0.2 and 0.4, respectively) thus suggesting a decrease in the cubic-to-tetragonal phase transition temperature with increasing $Eu^{2+}$.

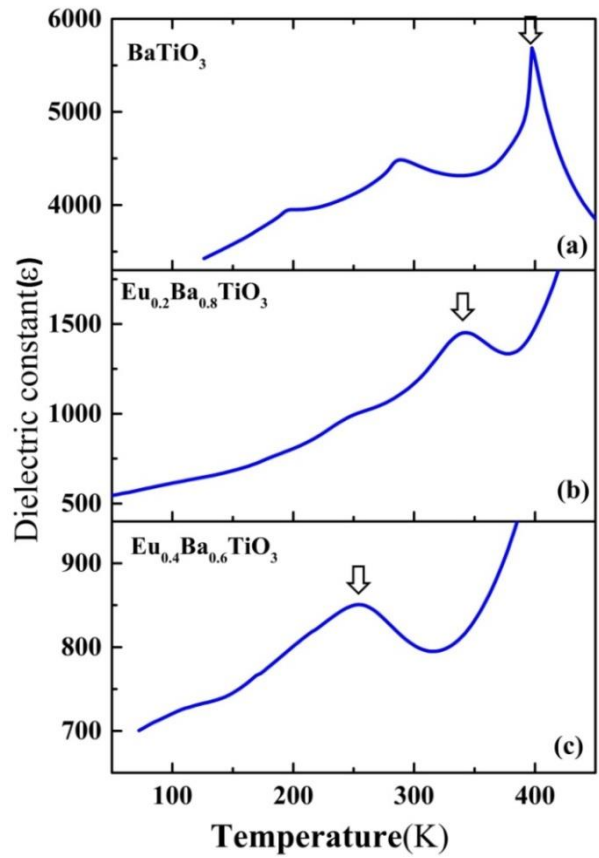

Figure S9: Dielectric constant as a function of temperature for (a) $BaTiO_3$, (b) $Eu_{0.2}Ba_{0.8}TiO_3$ and (c) $Eu_{0.4}Ba_{0.6}TiO_3$ measured at a frequency of 1 MHz thus corroborating the decrease in the tetragonal to cubic phase transition temperature, $T_C$, (indicated by arrows) with increasing content of Eu in $BaTiO_3$.